# Loc-Auth: Location-Enabled Authentication Through Attribute-Based Encryption

Marcos Portnoi    Chien-Chung Shen

Department of Computer and Information Sciences, University of Delaware, U.S.A.

{mportnoi,cshen}@udel.edu

*Abstract*—Traditional user authentication involves entering a username and password into a system. Strong authentication security demands, among other requirements, long, frequently hard-to-remember passwords. Two-factor authentication aids in the security, even though, as a side effect, might worsen user experience. We depict a mobile sign-on scheme that benefits from the dynamic relationship between a user's attributes, the service the user wishes to utilize, and location (where the user is, and what services are available there) as an authentication factor. We demonstrate our scheme employing Bluetooth Low Energy beacons for location awareness and the expressiveness of Attribute-Based Encryption to capture and leverage the described relationship. Bluetooth Low Energy beacons broadcast encrypted messages with encoded access policies. Within range of the beacons, a user with appropriate attributes is able to decrypt the broadcast message and obtain parameters that allow the user to perform a short or simplified login.

## I. INTRODUCTION

The typical sign-on or login procedure to a computer system involves a user entering a username and password by means of a physical or virtual keyboard. This authentication scheme functions by requesting a piece of information that only the computer and the user know. Its security relies on the difficulty, for an attacker, in obtaining or guessing the password. A password's "strength" is typically evaluated from the length of the password in bits (the longer the password, the more time it takes for an algorithm to "guess" it by iterating through all possible combinations, i.e., brute force) and how distant it is from usual combinations, such as dictionary words (usually first tested by password cracking algorithms based on dictionary attacks). This presents a first issue to the user: employing and remembering a sufficiently long, mixed-case, mixed-digit-letter-special character password that is considered strong.

As a second issue, in today's Internet-centric computer applications, a user is typically invited to register to dozens of websites/applications, each with its own requirements of username and password format and length. In an attempt to manage the myriad of usernames and passwords, users may be tempted to resort to weak passwords, which are easily broken by dictionary attacks or brute force, drastically decreasing the security of the authentication. A superior approach is utilizing password managers, programs that store usernames and passwords in an encrypted database, and can also generate long, random passwords. Obviously, the password manager itself needs a suitable strong master password. Additional security may be achieved by multiple-factor authentication, which involves entering a password and an extra, temporary information, which only the user and the computer know and it is shared through a secondary channel. Examples are keychain token authenticators, and the sending of temporary codes through email or SMS. The results are two-fold: stronger security, but more time and work required for user authentication and login.

A third issue emanates from the current's trend of consumer mobile devices that are typically devoid of physical keyboards, such as tablets and smartphones taking place of traditional desktop computers. Even when utilizing a password manager, typing a long, mixed-case, mixed-digit-and-letter password in diminutive screen keyboards is a cumbersome task and degrades user experience. The experience could be improved if the password is simplified (resulting, however, in a weak password), or by utilizing biometrics, such as the (yet underutilized) fingerprint reader in the mobile device.

The username/password authentication procedure may be viewed from two perspectives. The static one, wherein the login process and the system assume a user is at a pre-determined location and/or with a pre-determined device, from which the user signed on to the system. If the user moves to another location and/or switches the login device, then the previous sign-on is lost and a new sign-on must be performed. The mobile perspective is one wherein the user signs on from a mobile device and can usually remain logged in to the system if location is changed, provided a network connection is maintained while moving. While it may seem that, in the mobile version, location does not play an important part, location does provide a piece of information that we employ to leverage security.

In this paper, we explore the dynamic relationship among these three components: (1) location, i.e., where the user is and which services are available to the user in that location; (2) who the user is and what attributes the user possesses, i.e., the characteristics that can identify the user uniquely or as a member of specific groups; (3) which services the user wants to utilize. We contribute by employing this relationship as a secondary authentication factor that enables solutions for the issues presented earlier: (a) allows for a simplified yet secure user login, taking into consideration user's location, desired service, and user's credentials, improving the user experience when entering passwords; (b) enables a login session to "travel" with the user, such that devices in proximity and different services in the same location may "inherit" the login session as the user approaches. To capture the dynamic relationship, we construct

This work is supported in part by NSF DUE-1241711.

Loc-Auth, a Location-Enabled Authentication Service that employs the expressiveness of Ciphertext-Policy Attribute-Based Encryption (CP-ABE) [1] to encode access policies that are built on location, services for the user, and user attributes. We demonstrate a case study which relies on Bluetooth Low Energy or Bluetooth Smart beacons to construct indoor location information and access policy [2]. The user receives the transmission of beacons through a Bluetooth Low Energy-enabled device, such as the smartphone. Access attributes are not negotiated with the user at each time before the simplified sign-on procedure; the access rules and secure login are encoded directly in our usage of the encryption, contributing additional expressiveness and flexibility to the secondary authentication factor. If the user possesses attributes satisfying the access policy, the user will be able to acquire parameters to continue with the simplified sign-on procedure.

The paper proceeds in Section II to survey related work in using mobile devices as user authenticators, location sensors, and password strength. Section III presents our Location-Enabled Authentication Service. Section IV discusses our contributions and novel algorithms through a case study, and Section V analyzes the security of our design. Finally, Section VI concludes the paper with future directions.

## II. RELATED WORK

Using mobile devices as a means for access control is the topic of a number of initiatives. Near-Field Communication (NFC), a technology embedded in some smartphones and keycards, has been envisioned as an authentication method for payments and as a substitute to passwords [3]. The technology works at near range (a few centimeters) and allows for passive NFC tags, which harvest energy from the NFC transmitter's RF field. There are also designs proposing the use of cell phones as authenticators by employing their SIM cards (which have unique identifications), and as one-time-password generators [4, 5]. An evaluation of these techniques is presented in [6]. One biometric-based solution is Nymi, a bracelet that identifies the user's unique electrocardiogram pattern as an authenticator [7]. Upon confirming that the registered user is wearing the device, the bracelet communicates with a smartphone, and the related application would then authenticate the user to other devices via Bluetooth communication. Further security and operation details of this device are yet unavailable at the time of writing. ZoneIT [8] describes a prototype for controlling the functionality of mobile devices based on short-range radio communication systems. Base stations installed on premises exchange messages with users' mobile devices for authentication and defining the functionality that should be disabled in the area, such as ringtones in movie theaters, camera flashes in museums, etc. The authors analyze the security of the prototype against common attacks, and implementation and business issues.

Numerous research efforts have been conducted regarding indoor localization, which is typically not feasible by means of GPS. Auspicious research include localization by means of Wi-Fi [9] and Bluetooth [10]. These approaches generally involve having a mobile device discovering its current location utilizing information passed on from stationary objects, such as Wi-Fi Access Points and Bluetooth transmitters.

In [11], the authors discourse on the users' need of utilizing random, strong passwords without re-use. The authors claim that using and remembering multiple random, unique passwords is not only unfeasible, but also suboptimal, and password managers might signify a single point of failure and loss of cross-client portability. By utilizing objective functions, the authors contend that re-using simple passwords for services and sites that do not hold a user's valuable information is optimal, in the sense that it allows a user to dedicate brain memory to remember strong passwords utilized for sensitive sites. Our Location-Enabled Sign-On offers ways to mitigate these issues, by allowing the use of simpler passwords, but augmenting their security by capturing, with our algorithms and ABE, the convergence of location/user's attributes/services into an authentication factor.

## III. THE LOCATION-ENABLED AUTHENTICATION SERVICE

To enable and realize how location, available services, and user's attributes interrelate as an authentication factor, we build the Location-Enabled Authentication Service (Loc-Auth). This service, which acts as a trusted party, talks to backend systems or services and to the mobile devices, performing the necessary functions such that authentication is performed between the relying backend system/services, the mobile device, and the user. Loc-Auth is better viewed in a framework, composed of three layers: the Application Realm, the Location-Enabled Authentication Layer, and the Hardware Realm.

The *application realm* comprises applications and services that require authentication from a user, such as user's digital wallets, merchants' POS and website systems, and web browsers. These applications will utilize Loc-Auth through APIs, requesting and receiving authentication information. Here, we borrow OpenID's nomenclature: an application, service or website that relies on Loc-Auth for authentication information is referred as Relying Party (RP); the user or customer is referred as user or end-user.

The *Location-Enabled Authentication Layer* implements Loc-Auth, including its communication protocols, cryptosystems, key exchange, and a token authenticator algorithm (a function that generates a number $tk$ (the token) according to a seed and a message [12]). This layer is responsible for interfacing with the application and the hardware realms and providing multiple-factor authentication for the user and RPs through its key exchange based on the Attribute-Based Encryption (ABE) cryptosystem. In ABE, a set of "attributes" represents a user's set of credentials/attributes, and a formula with these attributes as input represents an access function. Instead of encrypting a ciphertext targeted to a specific secret key (as is done in conventional public key encryption), ABE produces ciphertexts and secret keys according to user attributes and access policies. In this work, we utilize Ciphertext-Policy ABE (CP-ABE), in which the attributes are coded into the secret keys held by individual users, and the formula over the attributes (access policy) is encrypted with the ciphertext [13]. Software agents act on behalf of applications from the application realm (both on RPs, and on users' mobile devices). The privacy module aims at controlling information sharing with the RPs. The main algorithms of the Location-Enabled Authentication Layer provide forms of secure key exchange and secure broadcast of access policies (using ABE at their core).

The *hardware realm* encompasses physical devices, such as Bluetooth Low Energy beacons, mobile devices, POS machines,

etc. This layer abstracts the physical devices as "hardware components." Through this abstraction, the Authentication Layer can request services from this layer, e.g., location information, without direct regard to the specific hardware utilized.

For the case study, we focus on indoor location; our method utilizes Bluetooth Low-Energy beacons and does not require that the mobile device knows, or reports, its present location. Loc-Auth infers the location of a mobile device by utilizing wireless-delimited broadcast zones, in which the beacons broadcast a CP-ABE encrypted message. If a mobile device is within that zone, and it is able to decrypt the broadcast message, then a key exchange is initiated between the mobile device and Loc-Auth, through the Bluetooth radios. As battery life is an enduring concern for mobile device users, Bluetooth Low Energy poses as a concrete technique to implement our zone beacons. The practicality of utilizing Bluetooth beacons installed throughout an indoor area has been positively demonstrated by large-scale deployments of Bluetooth-based beacons such as iBeacons from Apple. The range of the Bluetooth's radio transmission can also be used to assist in indoor location granularity or precision. Bluetooth class 2 radios, typically the ones used in mobile devices, have a theoretical maximum operating range of 10 m (for a maximum transmission power of 4 dBm); therefore, a mobile device is assumed to be within a beacon's range if it is within 10 m of the beacon.

## IV. CASE STUDY: THE LOCATION-ENABLED SIGN-ON

We explain the use of Loc-Auth via a case study of location-enabled sign-on, wherein end users acquire access to backend systems by enabling their location as additional authentication factor. Bluetooth Low Energy beacons are used to construct indoor location information and policy broadcast, and users possess a Bluetooth Low Energy device, such as a compatible smartphone, to receive the beacons. In this scenario, Bluetooth beacons, installed around an office space (as shown in Fig. 1), transmit an encrypted message containing a cryptographic session token. This token is encrypted using CP-ABE and encodes the access rules, through the predicate, that is desired for the range of that beacon. A user with a Bluetooth device captures the encrypted transmission and attempts to decrypt it by utilizing the user's ABE secret key. If the user has sufficient attributes to fulfill the encoded predicate in the ABE-encrypted message, then the user device will successfully decrypt the message and obtain the token.

The user is able to realize her location as authentication factor and perform a simplified, short-login to a computer or another device (such as a VoIP phone) within range of the Bluetooth beacon, by utilizing the user's smartphone to transmit a "location sign-on." The location sign-on procedure is built as follows. The office's backend system registers with Loc-Auth, negotiating and generating appropriate security parameters, desired attributes and access rules. These access rules for one specific floor could be, for instance, "employees of firm XYZ AND employees of the financial dept. AND clearance level > 3." Users of the office backend system register with Loc-Auth and receive security keys and attributes. Examples of attributes are "employee marketing dept.", "clearance level 2," and "intern." Loc-Auth runs Algorithm 1 to perform the encrypted message broadcast through the beacon. The interval between broadcasts should be configurable, aiming at striking a proper location accuracy and overhead balance. For example, beacon intervals in IEEE 802.11 are typically 100 TU (Time Units, where 1 TU = 1024 microseconds).

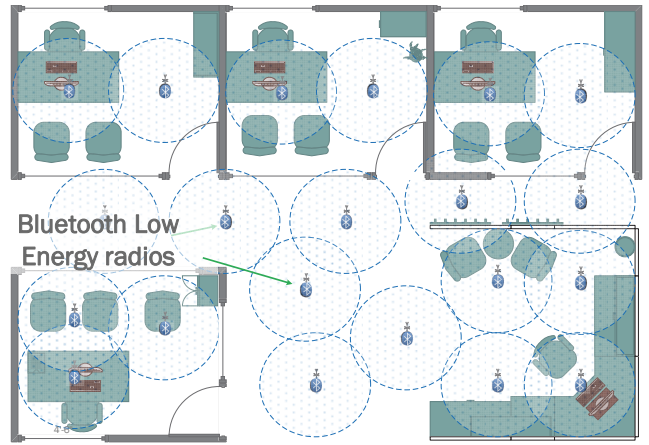

Fig. 1: Bluetooth Low Energy radios distributed around an office.

End-users run Algorithm 2 on their mobile devices. When those devices detect a beacon broadcast, they will attempt to decrypt the message, obtain data and collect additional authentication factors to perform the location sign-on, by transmitting the sign-on encrypted data back to the location beacons. The c-token is generated with the assistance of the token authenticator algorithm. Here, both the end-user and Loc-Auth share the individual user seed for the algorithm (exchanged during registration), and have synchronized clocks. Upon receiving a sign-on transmission, Loc-Auth runs Algorithm 3 and signs-on the end-user with the supplied authentication factors (if sufficient authentication was provided). To complete the access, the end-user may supply the final authentication factor, such as a simplified, short password to nearby terminals or devices. The backend system will, as a result, know about the end-user's current location. The signed-in session, as a common security procedure, should have an expiration time, after which Loc-Auth should request a new location authentication. A complete authentication with the location information only (without an extra factor such as the simplified password) is not always recommended, as any attacker with possession of the mobile device would then be able to sign-on on behalf of the user. By being smart about mobility, however, we may utilize this feature to make the location-enabled login session "travel" with the user between contiguous beacon cells through different RPs, as will be discussed later. In the case depicted, the beacon session token changes periodically and is only known by Loc-Auth. Both the user's mobile device and Loc-Auth, however, know the seed for the c-token, and this seed is unique to a user (thus both Loc-Auth and the user can generate the same c-token).

### A. Characteristics of Location-Enabled Sign-On

We may highlight the following characteristics about the authentication procedures performed by Loc-Auth, as compared with other traditional methods: (a) ABE is expressive, allowing access rules to be encoded in the message itself based on attributes. The message can then be broadcast through insecure medium. The primary access decision, then, need not rely on database access, or extended communication exchange between user and backend, or on specific users (but on "classes" of users).

Algorithm 1: Algorithm run by Location-Enabled Authentication Service for broadcast.

```
Repeat:
    Generates token for current time period;
    Encrypts token with desired access predicate
      using ABE;
    Broadcasts ABE(token);
end Repeat;
```

Algorithm 2: Algorithm run by end-user devices to detect beacon transmissions and perform location-enabled sign-on.

```
Listen to beacon broadcasts;
Receive ABE(token);
Decrypt ABE(token) with user's ABE private key;
if decryption is unsuccessful then
    goto End:;
else
    get token;
    Generate c-token for current time period;
    Hash token + password;
    Encrypt hash(token + password) using c-token
      as symmetric key;
    Encrypt username + c-token(hash(token +
      password)) using token as symmetric key;
    Send login=token[username +
      c-token(hash(token + password))];
end if;
End:
```

Algorithm 3: Algorithm run by Location-Enabled Authentication Service when receiving end-user sign-on.

```
Receive user's login;
Generate token for current time period;
Decrypt login and get username using token as
  symmetric key;
From username database, get user's seed and
  generate current user's c-token;
From username database, get user's password;
Generate hash(token + password);
Decrypt and get hash(token + password) from login
  using c-token;
Compare the locally generated hash(token +
  password) with one from login;
if there is a match then
    user is at this location; authenticate;
else
    do not authenticate user;
end if;
```

The decision is virtually transferred to the user, as a result of the user being able, or not, to decrypt the broadcast message. (b) Access rules can be changed on the fly simply by re-encrypting the new access predicate in the broadcast message. (c) The key broadcast using ABE is typically one-to-many, and not one-to-one as the traditional key exchanges. It should be noted, however, that some key exchange protocols, such as Diffie-Hellman's, provide key exchange for parties that have no prior knowledge of each other. The method described here requires that ABE private keys be distributed before through secure channels, such that a user can properly decrypt the broadcast messages if this user fulfills the predicate. Only beacons that broadcast a certain encrypted message accept the sign-on request from that decrypted message.

### B. Mobility and Hopping

If a user is outside of range of Bluetooth beacons, the interpretation is that Loc-Auth does not recognize the user's current location. It is then reasonable not to utilize that current location as an authentication factor in these situations. As mentioned before, the signed-in session has an expiration time, after which the location must be authenticated again. This procedure addresses beacon hopping, which happens as the user moves from one beacon area to another (thus, from one "location" to another). Hopping can, in addition, be used in other ways: the user agent, running in the mobile device, can periodically scan for beacon broadcasts even if the sign-on session is active and authenticated with the Service. In different RPs that allow for single sign-on, the user may be authenticated, in this manner, just by the act of moving within range of that second RP, coming from another RP. Hence, the sign-on "travels" with the user as she moves. For proper security, the location authentication keepalive time should not be long. In addition, Loc-Auth may employ a probabilistic analysis of the movement, such that the "traveling" or hopping does occur between neighboring areas, and not between eccentric, non-adjacent areas.

### V. SECURITY ANALYSIS

An attacker might utilize numerous vectors to attempt maliciously exploiting the mechanisms of Loc-Auth. We now analyze the behavior of Loc-Auth in the advent of three forms of attacks: replay attack, wormhole attack, and denial-of-service (DoS) attack. Our analysis for the first two cases is constructed as an interactive game with three players: Loc-Auth, a legitimate user, and a malicious attacker. The attacker attempts to gain access to the service, obtain information, or disrupt legitimate access.

### A. Replay Attack

**Setup**. Loc-Auth broadcasts encrypted message $m$ with token $tk$ at location $l$. An attacker at the same location listens to the medium and records $m$ at time $t$. Also, at this time, a legitimate user may listen to the broadcast message and proceed with the location sign-on. The attacker does not have knowledge of anything (i.e., secret keys, tokens) other than the original location $l$ of the broadcast and the time $t$, and is capable of either listening to or transmitting in the channel. The session token utilized to generate $m$ is only valid during a time $d$. At time $t + d$, Loc-Auth selects a new session token (and generates a new encrypted message), and henceforth message $m$ becomes invalid.

**Attack**. At time $t + d + \delta$, where $\delta > 0$, the attacker broadcasts the message $m$ at the same location $l$, such that legitimate users can listen to it.

**Response of the system**. A legitimate user listens to $m$. This user will run Algorithm 2 and, if possessing the required keys and attributes, will reply with the encrypted login (reply message $m_r$). Both Loc-Auth and the attacker will listen to the reply; the attacker, without knowledge of the secret keys, cannot decrypt the reply message $m_r$. In addition, the reply message for message $m$ at time $t$ is essentially distinguishable from the reply message $m_r$ for the same message $m$ now at time $t + d + \delta$, since the encryption algorithms used to generate the reply message are randomized; the attacker cannot infer information (e.g., cannot expect the same reply message from the same user, given the same message $m$ at any time).

Upon receiving $m_r$, Loc-Auth will run Algorithm 3. However, as the token generated by Loc-Auth at time $t + d + \delta$ is different from the token utilized in $m_r$ (and that was valid during $t + d$), authentication will fail. The results here are the same if, instead of recording and replaying $m$, the attacker records and replays $m_r$ at time $t + d + \delta$.

*B. Wormhole Attack*

**Setup**. Loc-Auth broadcasts encrypted message $m$ with token $tk$ at location $l$. An attacker at the same location listens to the medium and records $m$ at time $t$. As in the replay attack, the attacker has knowledge of the original location $l$ of the broadcast and the time $t$, but also has a direct network connection through a secondary channel to another location $p$. Normally, broadcasts in $l$ cannot be heard in $p$ and vice versa.

**Attack**. Within the time slot $t + d$ (wherein token $tk$ is still valid), the attacker broadcasts message $m$ at location $p$.

**Response of the system**. A legitimate user listens to $m$ and runs Algorithm 2; replies with the encrypted login (reply message $m_r$) if she has the proper ABE secret keys. Loc-Auth, receiving $m_r$, will run Algorithm 3. The token generated by Loc-Auth within time $t + d$ at location $p$ is different from the token $tk$ utilized in $m_r$, and that was generated at location $l$. Authentication will thus fail. Again, the results here are the same if, instead of recording and replaying $m$ from location $l$ to location $p$, the attacker records and replays $m_r$ from one location to another, within time $t + d$.

*C. Denial-of-Service Attack*

An attacker may perform a DoS attack by either (a) replaying previously recorded messages (broadcasts from the beacons, or reply messages from clients); (b) sending fake beacon or client messages; or (c) transmitting radio noise at the same frequencies as the Bluetooth beacons and clients, with such transmitting power and constancy as to disrupt legitimate communications between Loc-Auth and users. Bluetooth radios perform frequency hopping, in which the Bluetooth Master defines a hopping pattern, and then transmission will occur in different frequencies following the hopping pattern. This method mitigates the effects of radio jamming, since transmission can still occur in other frequencies that are not being jammed. Naturally, an applied attacker can still jam all possible Bluetooth frequencies. While it is technically complicated to resist such an attack, it is not problematic to uncover the attacker's radio transmitter and have it disabled [8, 14].

In the event that location sign-on fails completely, the backend system or relying party (RP) will fall back to a full authentication request, without the location authentication factor.

## VI. CONCLUSION AND FUTURE WORK

We proposed Loc-Auth, a location-enabled authentication service. By using Loc-Auth, we design a sign-on scheme that utilizes the dynamic relationship between a user's location, the available services at that location, and the user's attributes as an additional authentication factor and thus enables a simplified sign-on to be performed in devices or backend systems nearby. Additionally, the login session can "travel" with the user through different relying systems or devices as the user physically moves. Bluetooth Low-Energy beacons broadcast an encrypted message employing Ciphertext-Policy Attribute-Based Encryption (CP-ABE). A device in the user's possession, such as a smartphone, and within range of the beacons captures the broadcast message. If the user has appropriate attributes, the smartphone is able to decrypt the broadcast message and engages in a login process on behalf of the user. The relying backend system terminal or device in the immediate vicinity, then, may allow for a simplified login from that user or accept a "traveling" login session, granted that the user's location has been validated.

Initial experimentation focuses on implementing the authentication algorithm. By means of simulation, we test the effectiveness of the service and simplified login scheme when users are within range of different beacon areas. Our future work involves investigating the security of the proposed scheme according to security models. In particular, the task is to prove the algorithms to be (a) semantically secure, secure under (b) chosen ciphertext attacks (CCA) and under (c) chosen plaintext attacks (CPA). Moreover, we plan to investigate the issues of key revocation and update and analyze the performance of the model when subject to more attack vectors.